\documentclass{article}
\pdfoutput=1
\usepackage{cite}
\usepackage{amsmath}
\usepackage{amssymb}
\usepackage{dsfont}
\usepackage{ulem}
\usepackage{slashed}
\usepackage{multicol}
\usepackage{caption}
\usepackage[top=2.5cm,bottom=3.5cm,left=1.75cm,right=1.75cm]{geometry}
\usepackage{graphicx}
\usepackage{lettrine}

\newenvironment{Figure}
  {\par\medskip\noindent\minipage{\linewidth}}
  {\endminipage\par\medskip}

\usepackage{abstract}

\usepackage{titlesec}
\titleformat{\section}[block]{\large\scshape\centering}{\thesection.}{1em}{}
\titleformat{\subsection}[block]{\scshape}{\thesubsection.}{1em}{}

\title{\vspace{-15mm}%
	\fontsize{20pt}{10pt}\selectfont
	\textbf{Singlet Fermionic Dark Matter and the Electroweak Phase Transition}
	}	
\author{%
	\large
	\textsc{Malcolm Fairbairn\footnote{malcolm.fairbairn@kcl.ac.uk} and Robert Hogan\footnote{robert.hogan@kcl.ac.uk}}
	\\[2mm]
	\normalsize	Physics Department, \\
	\normalsize	King's College London \\
	\vspace{-5mm}
	}
\date{}

\begin{document}
\normalem
\maketitle

\begin{abstract}
\noindent We consider a model with a gauge singlet Dirac fermion as a cold dark matter candidate. The dark matter particle communicates with the Standard Model via a gauge singlet scalar mediator that couples to the Higgs. The scalar mediator also serves to create a tree-level barrier in the scalar potential which leads to a strongly first order electroweak phase transition as required for Electroweak Baryogenesis. We find a large number of models that can account for all the dark matter and provide a strong phase transition while avoiding constraints from dark matter direct detection, electroweak precision data, and the latest Higgs data from the LHC. The next generation of direct detection experiments could rule out a large region of the parameter space but can be evaded in some regions when the Higgs-singlet mixing is very small.
\end{abstract}

\begin{multicols}{2}

\section{Introduction}
The Standard Model (SM) of particle physics is a remarkably successful theory. The recent discovery of a 125 GeV (Brout-Englert-Kibble-Guralnik-Hagen-)Higgs boson by the ATLAS\cite{ATLAS1207} and CMS\cite{CMS1207} collaborations at the LHC has confirmed the existence of the last piece of the puzzle. It has long been known however that the SM has many shortcomings. Some of these, such as the much discussed hierarchy problem, are theoretical problems which do not conflict with any experimental measurements. Perhaps more importantly the SM is incapable of explaining some well established observational results. Two of the most notable of these are the presence of a large non-baryonic dark matter (DM) component of the energy density in the Universe and the asymmetry between matter and anti-matter.

Perhaps the most common solution to the DM problem uses an R-parity conserving supersymmetric framework where the lightest supersymmetric particle plays the role of the DM. The absence so far of any evidence for supersymmetry at the LHC makes it difficult to understand top-down expectations and perhaps motivates a more general bottom-up approach. The simplest such solution is to consider an additional gauge singlet scalar degree of freedom, $s$, that couples only to the Higgs boson \cite{McDonald0702,Davoudiasl0405,Burgess0011,O'Connell0611,BahatTreidel0611, Barger0706,He0701,Gonderinger1202}. To ensure $s$ is stable a $\mathds{Z}_2$ symmetry, $s \rightarrow -s$, is imposed. The next simplest model is to introduce two gauge singlets, one fermion and one scalar, and to let the fermion act as the DM particle \cite{Kim0611,Kim0803,Qin11,LopezHonorez1203,Baek1209,Farina1303,Petraki0711}. Such models, in which the DM only couples to the SM particles via the Higgs, are sometimes known as Higgs Portal DM.

In order to obtain a baryon asymmetry in the Universe a model must satisfy the Sakharov conditions \cite{Sakharov1967}: (i) baryon number violation, (ii) $C$ and $CP$ violation, and (iii) non-equilibrium dynamics. Sphaleron processes and the $CP$ odd phase of the CKM matrix provided solutions to conditions (i) and (ii) (although it is now known that there is insufficient $CP$ violation in the SM \cite{Gavela9312,Gavela9406,Huet9404}).  Electroweak Baryogenesis \cite{Kuzmin85} attempts to use the electroweak phase transition to  fulfil (iii)  and establish baryon asymmetry dynamically (see \emph{e.g.}\cite{Cohen9302,Trodden9803,Morrissey1206} for reviews). For this to work the electroweak phase transition needs to be strongly first order to provide sufficiently out of equilibrium dynamics and prevent washout of the asymmetry by sphaleron processes below the electroweak scale. Unfortunately, the measurement of the Higgs mass means that the phase transition in the Standard Model is not strongly first order and baryogenesis isn't viable via this route. 

There have been many attempts to create a first order electroweak phase transition by extending the Standard Model. There has been considerable focus in the literature \cite{Pietroni9207,Menon0404,Profumo0705,Cline0905,Espinosa1107, Chung1209,Cline1210,Patel1212} on the need for tree-level barriers in the electroweak potential rather than appealing to large thermal loop corrections. This can be readily achieved by including extra scalar degrees of freedom (\emph{e.g.} \cite{Espinosa9301, Chung1209,Choi9308,Ham0411, Ahriche0701,Profumo0705,Cline0905,Espinosa1107,Espinosa1110,Cline1210,Patel1212}).

It is therefore natural to ask if these minimal models can solve both the DM and baryon asymmetry problems at the same time. Many attempts were made in this direction using scalar extensions \cite{Barger0706, Barger0811,Chowdhury1110,Ahriche1201,Cline1210,Cline1302} and it was found in the case of the real scaler singlet it was impossible to solve both problems together. Recently for example, it was found \cite{Cline1210} that the addition of $\mathds{Z}_2$ singlet cannot simultaneously provide a strong electroweak phase transition and account for all of the DM. This was due to a conflict between the requirement of a large barrier and direct detection constraints that both depend strongly on one parameter: the Higgs-scalar coupling (although this conflict might be evaded in composite scalar scenarios with more complicated hidden sectors \cite{Frigerio1204}). In this paper we propose a simultaneous solution of the total DM relic density and a first order phase transition by relaxing the need for a $\mathds{Z}_2$ symmetry for the scalar. The scalar is therefore no longer stable and we introduce a fermionic singlet to act as the DM. 
\section{The Model}
The most general renormalizable tree-level potential of the SM+scalar singlet is given by
\begin{equation}
\begin{split}
V=&-\frac{1}{2}u_h^2 h^2+\frac{1}{4}\lambda_h h^4 +\frac{1}{2}u_s^2 s^2+ \frac{1}{4}\lambda_s s^4 \\
&+\frac{1}{4}\lambda_{hs} s^2 h^2+\mu_1^3 s+\frac{1}{3} \mu_3 s^3 +\frac{1}{4} \mu_m s h^2,
\end{split}
\end{equation}
where $h$ and $s$ are the physical Higgs and singlet fields respectively. The final three terms are obtained by relinquishing the requirement of a $\mathds{Z}_2$ symmetry for the $s$ field. This potential is invariant under shifts of the singlet field, $s \rightarrow s+\delta$, which amounts to a redefinition of parameters. We will use this freedom to set $\mu_1=0$ throughout.

The singlet fermion enters through the Lagrangian
\begin{equation}
\mathcal{L}_{DM}=\bar{\psi} (i \slashed \partial -m )\psi+ g_s s \bar{\psi} \psi.
\end{equation}
Since $s$ will in general attain a vacuum expectation value (vev), $\langle s \rangle = w$, the dark matter mass is $m_{\psi}=m+g_s w$. We will take $m_{\psi}$ to be a free parameter because $m$ may be chosen freely. To prevent $\psi$ from mixing with the SM fermions it must carry a global $U(1)$ fermion number symmetry. In the absence of this quantum number it would couple to the SM like a right-handed neutrino and would be unstable.

For a general choice of parameters both $h$ and $s$ will attain vevs. This will introduce some mixing of the gauge eigenstates $h$ and $s$. To describe physical process we have to transform into the mass eigenbasis. The mass eigenstates are given by
\begin{align*}
h_1 &=\sin \alpha  \ s+\cos \alpha \  h \\
h_2 &=\cos \alpha \ s -\sin \alpha \  h,
\end{align*}
where the mixing angle $\alpha$ is defined by
\begin{equation}
\tan \alpha =\frac{x}{1+\sqrt{1+x^2}}
\end{equation}
with 
\begin{equation}
x=\frac{2 m_{sh}^2}{m_h^2-m_s^2}
\end{equation}
and
\begin{equation}
m_h^2=\left. \frac{\partial^2 V}{ \partial h^2} \right|_{(v,w)}\ m_s^2= \left. \frac{\partial^2 V}{ \partial s^2} \right|_{(v,w)}\ m_{s h}^2= \left. \frac{\partial^2 V}{ \partial h \partial s} \right|_{(v,w)} ,
\end{equation}
where $v=\langle h \rangle= 246$ GeV is the Higgs vev. The mass eigenvalues are given by
\begin{equation}
m_{h_{1,2}}^2 =\frac{1}{2}(m_h^2+m_s^2) \pm \frac{m_{sh}^2}{x}\sqrt{1+x^2}.
\end{equation}
The definition of $\alpha$ ensures that $\cos \alpha > \frac{1}{\sqrt{2}}$ so $h_1$ is identified as the Higgs-like state and $h_2$ as the singlet-like state. This mixing will introduce a coupling both between $h_1$ and $\psi$, and between $h_2$ and the SM particles and so will modify the Higgs phenomenology of the SM as we will now discuss.

\subsection{Higgs Physics Constraints}
Throughout the analysis we take the Higgs-like scalar mass to lie in the $95\%$ confidence interval \cite{ATLAS-CONF-2012-163},
\begin{equation}
m_{h_1} =125.2 \pm 1.8\ \text{GeV}.
\end{equation}
The search for a SM Higgs-like boson at the LHC and the subsequent measurement of its properties also provides constraints for any model with an extended Higgs sector. The mixing of the Higgs and singlet states introduces a universal $\cos \alpha$ suppression of all couplings of the Higgs-like particle, $h_1$, relative to the SM Higgs. The measured signal strengths of the 125 GeV Higgs boson will therefore constrain the allowed values of $\cos \alpha$. 

Additionally, the possibility of non-SM decays of the Higgs will also introduce a universal suppression, because it will dilute the branching ratios to all SM final states. If kinematically allowed, this model provides new decay modes for the Higgs-like particle $h_1$: $h_1 \rightarrow 2h_2,\  \bar{\psi} \psi $. There is a degeneracy between a universal suppression factor and additional Higgs decay modes that cannot be lifted until any additional branching ratio is measured directly \cite{Espinosa1205}. We therefore absorb the non-standard branching ratio into a redefined suppression factor. In the presence of both effects the signal strength, $\mu$, will be modified according to 
\begin{equation}
\mu=\cos^2 \alpha \left(1-BR^1_{BSM}\right) \mu_{SM} =a'^2 \mu_{SM},
\end{equation}
where $BR^1_{BSM}$ is the branching ratio of $h_1$ to non-standard model final states. Recent global analyses of Higgs couplings \cite{Ellis1303,Falkowski1303} found that such a suppression is required to have $a' > 0.9$ at 95\% CL in order to account for the observed Higgs signal strengths.  We will see later that this constraint can be easily satisfied while satisfying both DM and electroweak phase transition constraints. 

The non-observation of an additional Higgs boson can also provide a constraint on our model. The exclusion plots produced by the ATLAS and CMS collaborations are however based on a particle with SM Higgs-like couplings. In our case, all couplings of the singlet-like state will be suppressed by a factor of $\sin \alpha$ relative to the Standard Model Higgs coupling. By analogy with Higgs-like state above, the signal strength, $\mu$, will be modified to
\begin{equation}
\mu=\sin^2\alpha (1-BR^2_{BSM})\mu_{SM} = b'^2 \mu_{SM},
\end{equation}
where $BR^2_{BSM}$ includes decays to $\psi \bar{\psi}$ and $2h_1$. The current best exclusion data\cite{CMS1304} sets a conservative bound of $b'^2 \lesssim 0.1$ for $m_{h_2} \lesssim 400$ GeV but the bound becomes significantly weaker for larger scalar masses. This is again easily avoided in our model.

We also require our potential, $V$, to be well behaved. That is, we require $V$ to be absolutely stable at tree-level by ensuring that it does not develop any directions which are unbounded from below. We do not, however, require stability at higher energy scales. We also ensure that electroweak symmetry breaking is viable, and that the broken vacuum state we end up in is the global minimum of the zero temperature potential. We insist that the magnitude of all dimensionless couplings are reasonably small ($< 1.5$) so that the theory will be perturbative, although a full analysis of the relevant renormalization group equations has not be carried out.

Finally, we ensure constraints coming for electroweak precision observables are avoided using \cite{Baek1112,Baak1209,Eberhardt1209}. We find, however, that this does not significantly add to the constraints on the model once other constraints are met.
\section{Dark Matter}
We will now consider the constraints coming from the requirement that the gauge singlet fermion, $\psi$, plays the role of the cold dark matter of our universe. 
\subsection{Relic Density}
The recent data from the Planck satellite \cite{PlanckXVI} have provided the strongest constraints to date on cosmological parameters. The 95\% CL on the physical DM relic density is
\begin{equation}
0.1134< \Omega_{DM} h^2 < 0.1258,
\end{equation}
where $h \simeq 0.7$ is the Hubble constant in units of $100$ km s$^{-1}$ Mpc$^{-1}$. 

The relic density of the $\psi$ field is produced via the usual process of thermal freeze-out. The annihilation of $\psi \bar{\psi}$ proceeds via $h_i$ mediated $s-$channel diagrams $\psi \bar{\psi} \rightarrow f \bar{f},\ W^+W^-,\ ZZ,\ h_i h_j,\ h_i h_j h_k$, with $i,j,k=1,2$. We also consider $t-$ and $u-$channel annihilation into $h_i h_j$ final states. We consider only the dominant contribution of $b\bar{b}$ and $t\bar{t}$ to the two fermion final state and neglect other subdominate channels. The cross sections to 3 body final states were calculated but it was found they were highly suppressed by the three body phase space and so are neglected from the final analysis. The full expression of $\sigma v$ and the relevant couplings are contained in the Appendix.

Rather than using a velocity expansion to approximate the thermally averaged cross section we adopt the approach of \cite{Gondolo1990} and preform the thermal averaging explicitly. This approach is more reliable than a velocity expansion in regions near resonances and thresholds which are often crucial. The thermally averaged cross section at temperature, $T$, is then given by 
\begin{equation}
\langle \sigma v_{rel} \rangle = \frac{1}{8m_{\psi}^4 T K_2^2(\frac{m_{\psi}}{T})}
 \int_{4m_{\psi}^2}^{\infty} ds\ \sigma\ s^{3/2} \beta_{\psi}  K_1\left(\frac{\sqrt{s}}{T}\right),
\end{equation}
where $\beta_{\psi}=\sqrt{1-4m_{\psi}^4/s}$, $\sqrt{s}$ is the centre of mass energy, and $K_{1,2}$ are the modified Bessel functions.

The standard approximate solution to the Boltzman equation relates the relic density the thermally averaged annihilation cross section via
\begin{equation}
\Omega_{DM} h^2 \simeq \frac{(1.07 \times 10^9) x_F}{\sqrt{g_{\star}}M_p \langle \sigma v_{rel} \rangle},
\end{equation}
where $x_F=m_{\psi}/T_{F}$, $T_F$ is the freeze-out temperature, $M_p$ is the Planck mass, $g_{\star}$ is the effective number of relativistic degrees of freedom in thermal equilibrium at $T_F$, and all quantities are in GeV. The freeze-out temperature is solved for iteratively using 
\begin{equation}
x_F =\log\left( \frac{m_{\psi}}{2 \pi^3} \sqrt{\frac{45 M_p^2}{2 g_{\star} x_F}}\langle \sigma v_{rel} \rangle \right).
\end{equation}

The fulfilment of the relic density condition is highly constraining and permits only a narrow band in the $(m_{\psi},g_s)$ plane. Figure \ref{fig: relic dense} shows the result for two choices of singlet-like mass, $m_{h_2}$. Clear features can be seen in the Figure. Notable are the $s$-channel resonances for both $h_1$ and $h_2$ and the thresholds corresponding the $WW$ and $h_i h_i$ channels opening. Once all channels are open the $(m_{\psi},g_s)$ plane becomes even more constrained and the coupling $g_s$ tends to 1 for $m_{\psi}=1$ TeV.
\begin{Figure}
\centering
\includegraphics[width=\linewidth, height=.6\linewidth]{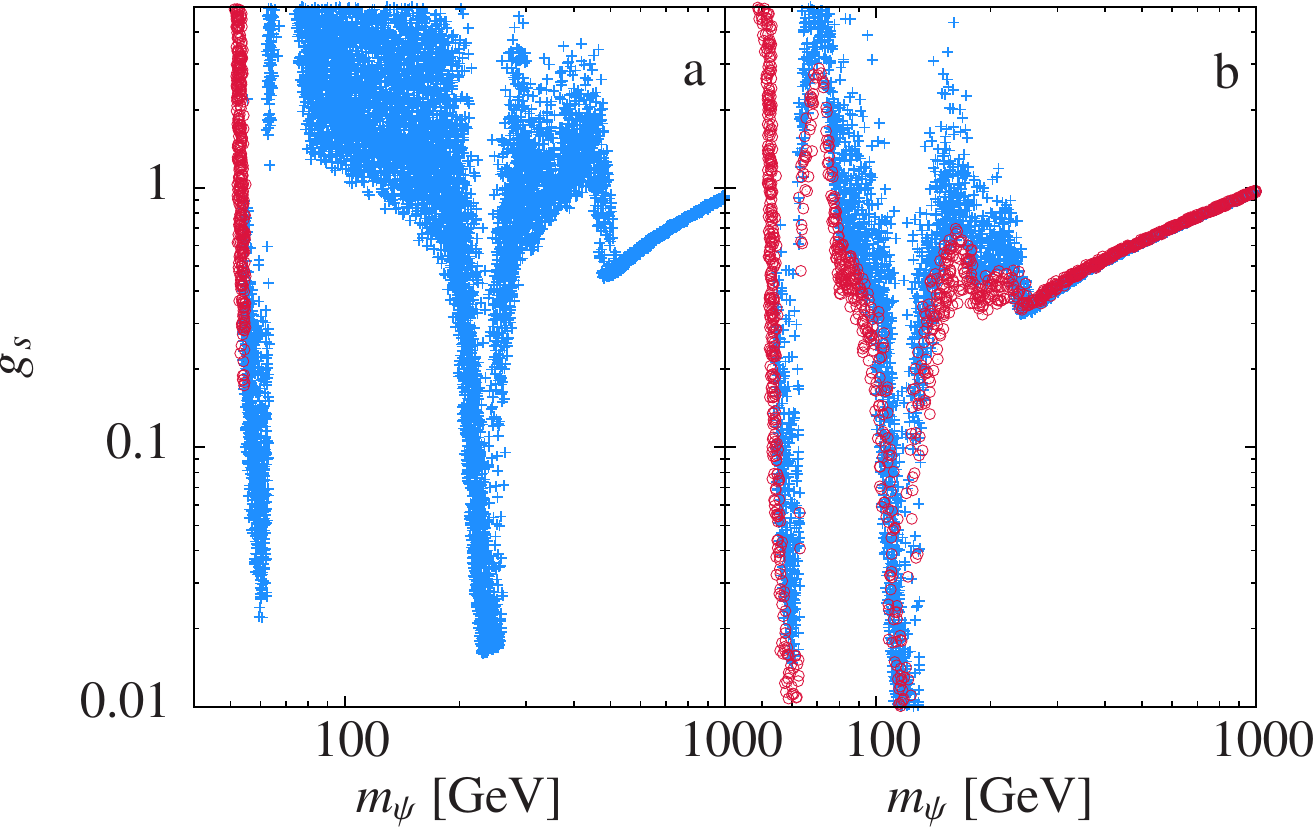}
\captionof{figure}{\it Points in $(g_s,m_\psi)$-plane satisfying Planck relic density constraints for (a) $m_{h_2}=500$ GeV $(\pm 5\%)$, and (b) $m_{h_2}=250$ GeV $ (\pm5\%)$. The red points are ruled out by LHC Higgs physics.}
\label{fig: relic dense}
\end{Figure}

We note that Sommerfield enhancement \cite{Sommerfield31} can play a role \cite{LopezHonorez1203} but it will not change the qualitative features of our result and so is neglected.

\subsection{Direct Detection}
There has been substantial effort to detect the presence of WIMP dark matter directly through elastic scattering off heavy nuclei. The effective WIMP-nucleon couplings depend strongly on the details of the nuclear physics, and are given by \cite{Nihei0404,Ellis0001}
\begin{equation}
f_{p,n}=m_{p,n} \bar{\alpha}\left( \displaystyle\sum\limits_{q=u,d,s} f_{Tq}^{p,n} +\frac{2}{9}f_{Tg}^{p,n}\right),
\end{equation} 
where the hadronic matrix elements, $f_{Tq}^{p,n}$, are given in \cite{Ellis0001}, and $f_{Tg}^{p,n}=1- \sum f_{Tq}^{p,n}$. Here $\bar{\alpha}$ is related to the model-dependent coupling of the dark matter to quarks, $\alpha_q$, by \cite{Kim0803}
\begin{equation}
\bar{\alpha}=\frac{\alpha_q}{m_q}=\frac{g_s \cos \alpha \sin \alpha}{v}\left(\frac{1}{m_{h_1}^2} -\frac{1}{m_{h_2}^2}\right),
\end{equation}
where the scattering proceeds via t-channel exchange of $h_{1,2}$.  The cross section for WIMP-nucleus scattering is then
\begin{equation}
\sigma_N=\frac{4 M_r^2}{\pi}\left(Z f_p +(A-Z)f_n \right)^2,
\end{equation}
where $1/M_r=1/m_{\psi} +1/m_{N}$ is the reduced mass of the system. For ease of comparison with experiment this is translated into the spin-independent cross section per nucleon via \cite{Bottino9612}
\begin{equation}
\sigma_{SI}=\frac{1+m_{\psi}^2/m_N^2}{1+m_{\psi}^2/m_p^2} \frac{\sigma_N	}{A^2}.
\end{equation}

The current best limits on $\sigma_{SI}$ are provided by the Xenon100 experiment \cite{Xenon1207} so we use this data to constrain our model. We see in Figure \ref{fig:direct detection} that when $m_{\psi} < \frac{1}{2} m_{h_2}$ we rely on resonant s-channel annihilation in order to satisfy the relic density constraints while evading direct detection. When $m_{\psi} > \frac{1}{2} m_{h_2}$ however we see that a large proportion of the parameter space is below the Xenon bound. 

\begin{Figure}
\centering
\includegraphics[width=\linewidth ,height=.6\linewidth]{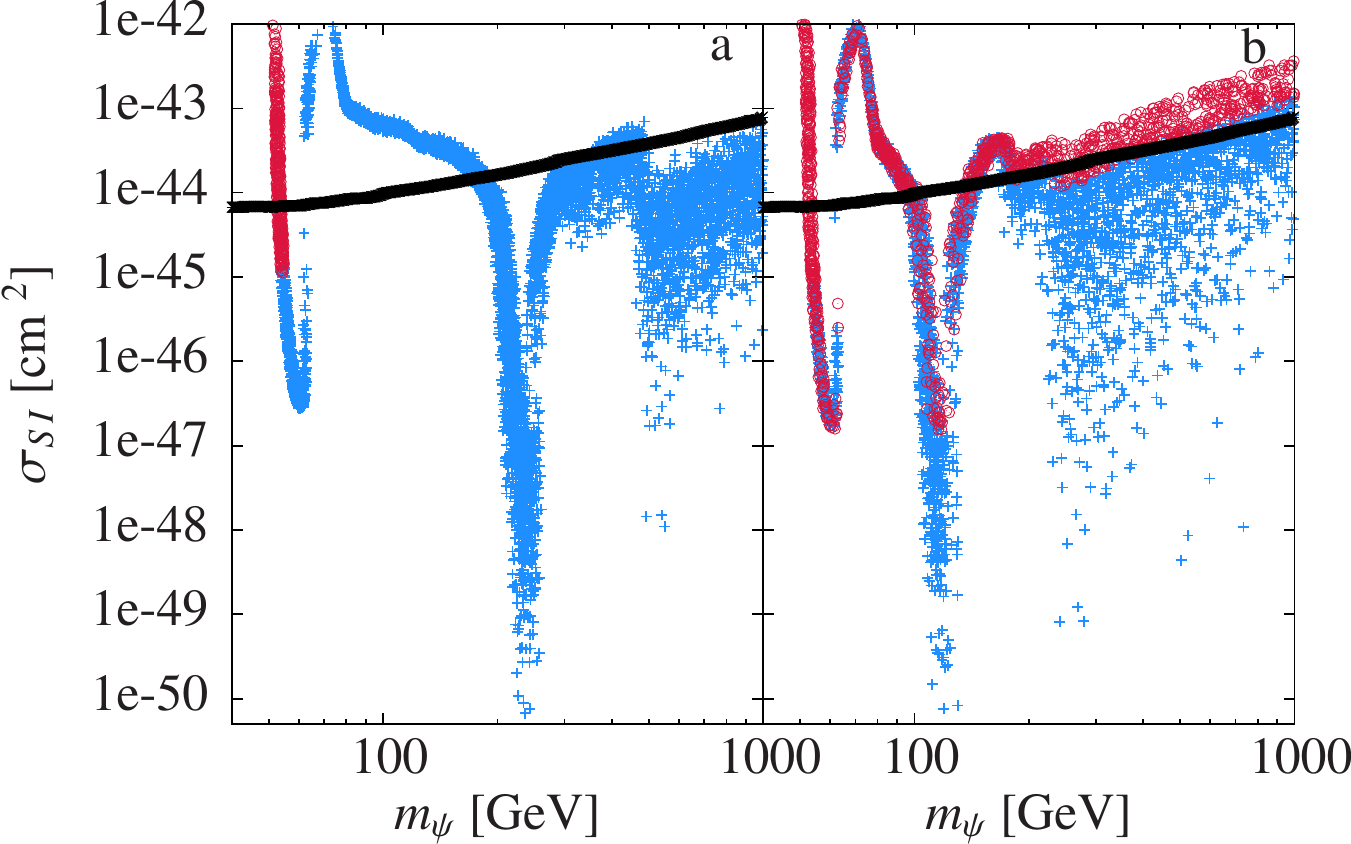}
\captionof{figure}{\it The scattering cross section per nucleon satisfying Planck relic density constraints for (a) $m_{h_2}=500$ GeV $(\pm 5\%)$, and (b) $m_{h_2}=250$ GeV $ (\pm5\%)$. The red points are ruled out by LHC Higgs physics. The Xenon100 2012 bound is included for comparision.}
\label{fig: relic dense}
\label{fig:direct detection}
\end{Figure}
It is also worth noting that the $h_2$ resonance reaches $\sigma_{SI} \sim 10^{-51}$ cm$^2$, which would evade detection by the next generation of direct detection experiments (e.g. \cite{Xenon1T}). It was found in \cite{LopezHonorez1203} that for annihilation diagrams that are independent of $\sin \alpha$, $g_s$ is constrained by relic density considerations but $\sigma_{SI}$, which is determined by $ g_s \cos \alpha \sin \alpha$, can be very small because $\sin \alpha$ remains unconstrained. This is the case  for $t$- and $u$-channel diagrams $\psi \bar{\psi} \rightarrow h_2 h_2$ and also for the $s$-channel diagram $\psi \bar{\psi} \rightarrow h_1 h_1$ (see Appendix for the relevant coupling) which becomes important at the $h_2$ resonance.

We note here that there is currently significant error in the values of the hadronic matrix elements. The points in Figure \ref{fig:direct detection} are quite sensitive to these errors but for simplicity we take central values and note that this may be subject to change.

\section{Electroweak Phase Transition}

The electroweak phase transition is the transition from $\langle h \rangle =0$ to $\langle h \rangle = v$. In the case where there are two scalar fields the vev of the second field may also change. The transition therefore proceeds via $(\langle h \rangle,\langle s \rangle)=(0,w_0) \rightarrow (v,w)$, where $w_0$ is not necessarily zero (although it is always possible to make this choice by shifting $s$). In order for a phase transition to be considered strongly first order we must have $\bar{v}(T_c)/T_c >1$, where $T_c$ is the critical temperature for the phase transition and $\bar{v}(T)$ is a gauge-independent, temperature-dependent, scale that characterises the sphaleron energy \cite{Carson90,Patel1101}. In the high-$T$ effective theory considered here this scale coincides with $v_c$, the vev of the Higgs field at the critical temperature, although this is not true of the full 1-loop effective potential. Also, the use of the high-$T$ effective theory means the gauge-dependence issues in the determination of $T_c$ discussed in \cite{Patel1101,Wainwright1204} do not apply because the problematic terms with linear and non-analytic $T$-dependence are dropped.

Traditionally, attempts to create a strongly first order phase transition in extensions of the SM have relied on large couplings of the Higgs to new bosonic degrees of freedom. This leads to significant thermal corrections that can render the phase transition first order. Unfortunately these contributions are proportional to the temperature. Since the important ratio is $\bar{v}/T_c$, the temperature dependent contributions to $\bar{v}$ are largely cancelled by $T_c$ and it is difficult to make $\bar{v}/T_c$ large. With the presence of barrier in the scalar potential at zero temperature it is possible to get strong first order phase transitions with only small thermal corrections. This means that $\bar{v}$ is much less temperature sensitive and $\bar{v}/T_c$ can be made very large by lowering $T_c$. The smallness of the loop corrections means it is only necessary to retain the leading order terms in the high-$T$ expansion of the one-loop thermal potential
\begin{equation}
V_T=\left( \frac{1}{2}c_h h^2 +\frac{1}{2} c_s s^2 +m_3 s\right) T^2,
\end{equation}
where
\begin{align}
c_h &= \frac{1}{48}\left( 9g^2 +3g'^2 +12y_t^2+24 \lambda_h+2\lambda_{hs}\right),\nonumber\\
c_s &= \frac{1}{12}\left( 2 \lambda_{hs} +3 \lambda_s +g_s^2\right), \\
m_3 &=\frac{1}{12}\left( \mu_3 +\mu_m \right)\nonumber.
\end{align}
Here, $g$ and $g'$ are the electroweak gauge couplings and $y_t$ is the top quark Yukawa coupling (the contributions of the other quarks are sub-dominant and so have been neglected). In the $T \to \infty$ limit these thermal contributions drive $(\langle h \rangle,\langle s \rangle ) \rightarrow (0,-m_3/c_s)$. The high-$T$ expansion can only be trusted up to $\bar{v}/T_c \sim 4$ so we do not consider values larger than this. It is also possible that if $\bar{v}/T_c$ becomes too large that tunnelling probability becomes too small for the phase transition to take place during the age of the universe \cite{Espinosa0809}. This typically occurs in the range $\bar{v}/T_c \sim 3-4$ but is model dependent, and removing these cases has no qualitative effect on our results so we omit this from the analysis.

In order to search numerically for models with $\bar{v}/T_c >1$ we closely follow the recipe provided in \cite{Espinosa1107}. We find that first order phase transitions mostly fall into two broad classes: a) $\Delta w=w-w_0 < 0$  and b) $ \Delta w >  0$ (note that the sign is coordinate basis dependent and can always be swapped by relabelling $(s,\mu_m,\mu_1, \mu_3) \rightarrow (-s,-\mu_m,-\mu_1, -\mu_3)$ but the relative sign between the two modes of breaking will remain). Figure \ref{ewptfig} shows the shape of the thermal potential at $T=0,T_c$ for each case. We note here that in the case of a $\mathds{Z}_2$ model it was pointed out in \cite{Espinosa1107} that for case b) it is not possible to have tree-level barrier, and that the case (a) must proceed via $(0,w_0)\rightarrow (v,0)$. If the vacuum had non-zero $w$ the $\mathds{Z}_2$ symmetry is broken by the vacuum and renders the scalar unstable.

\begin{Figure}
\centering
\includegraphics[width=\linewidth]{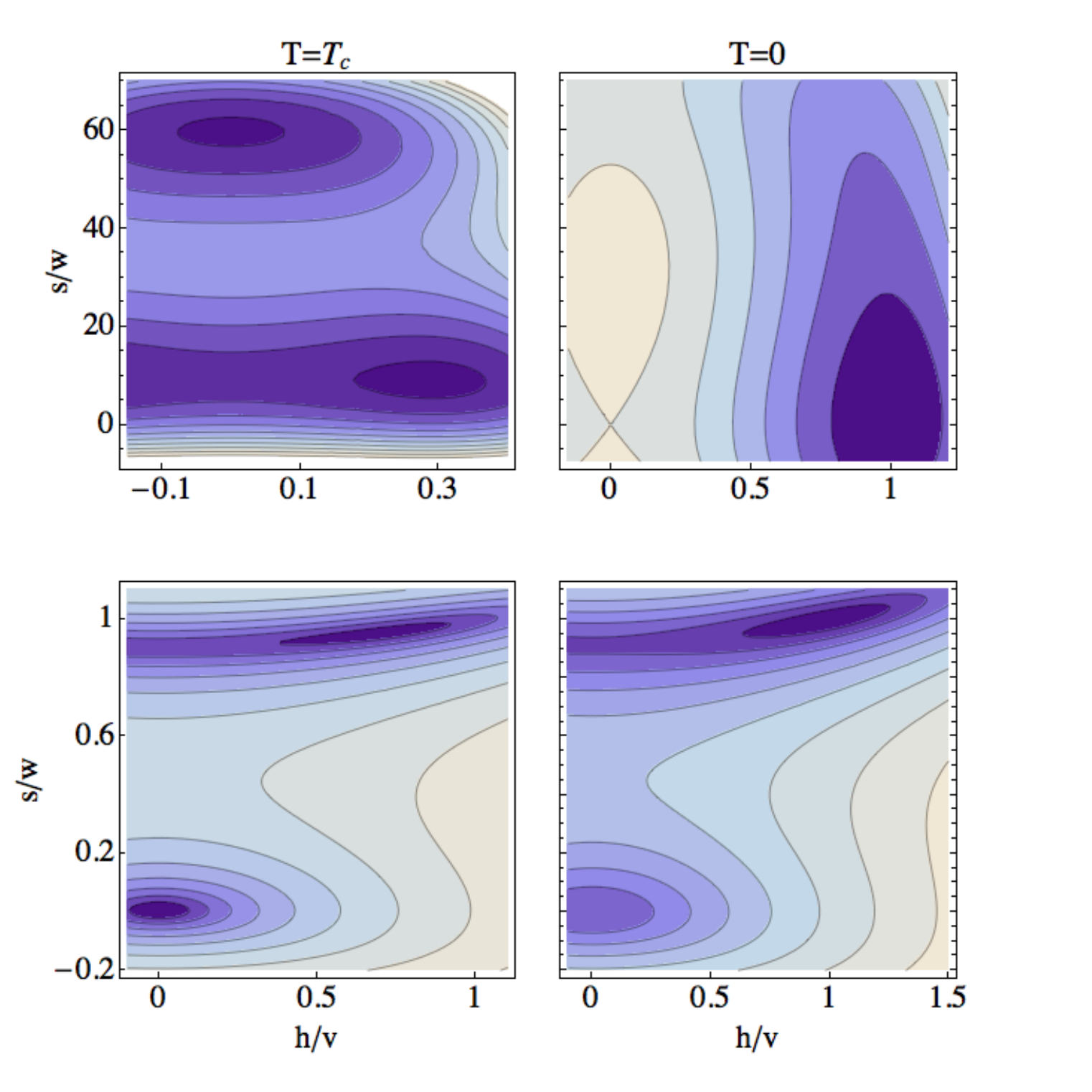}
\captionof{figure}{\it Thermal effective potential at $T=0,\ T_c$ for $\Delta w < 0$ (top), and $\Delta w > 0$ (bottom). The potential at $T=T_c$ shows two degenerate minima and as $T$ is lowered the electroweak breaking vacuum becomes the global minimum.}\label{ewptfig}
\end{Figure}

After a large monte carlo scan\footnote{We use flat priors for a convenient choice parameters described in \cite{Espinosa1107}. These parameters are then transformed into those described in this paper. As such, the density of points in plots (e.g. Figure \ref{fig: relic dense}) does not necessarily correspond directly to regions of higher probability.} of the parameter space we found many models that escape all constraints. Figure \ref{fig:vc_Tc} shows the distribution of the derived values of $\bar{v}/T_c$. We see that we can easily achieve large values for $\bar{v}/T_c$. Figure \ref{fig: params} shows the distribution of some of the key parameters after all cuts are made. We see that Higgs physics constraints, summarised in $a'$ and $b'$, are generically avoided in models that satisfy the DM and phase transition constraints. 

\begin{Figure}
\centering
\includegraphics[width=.6\linewidth]{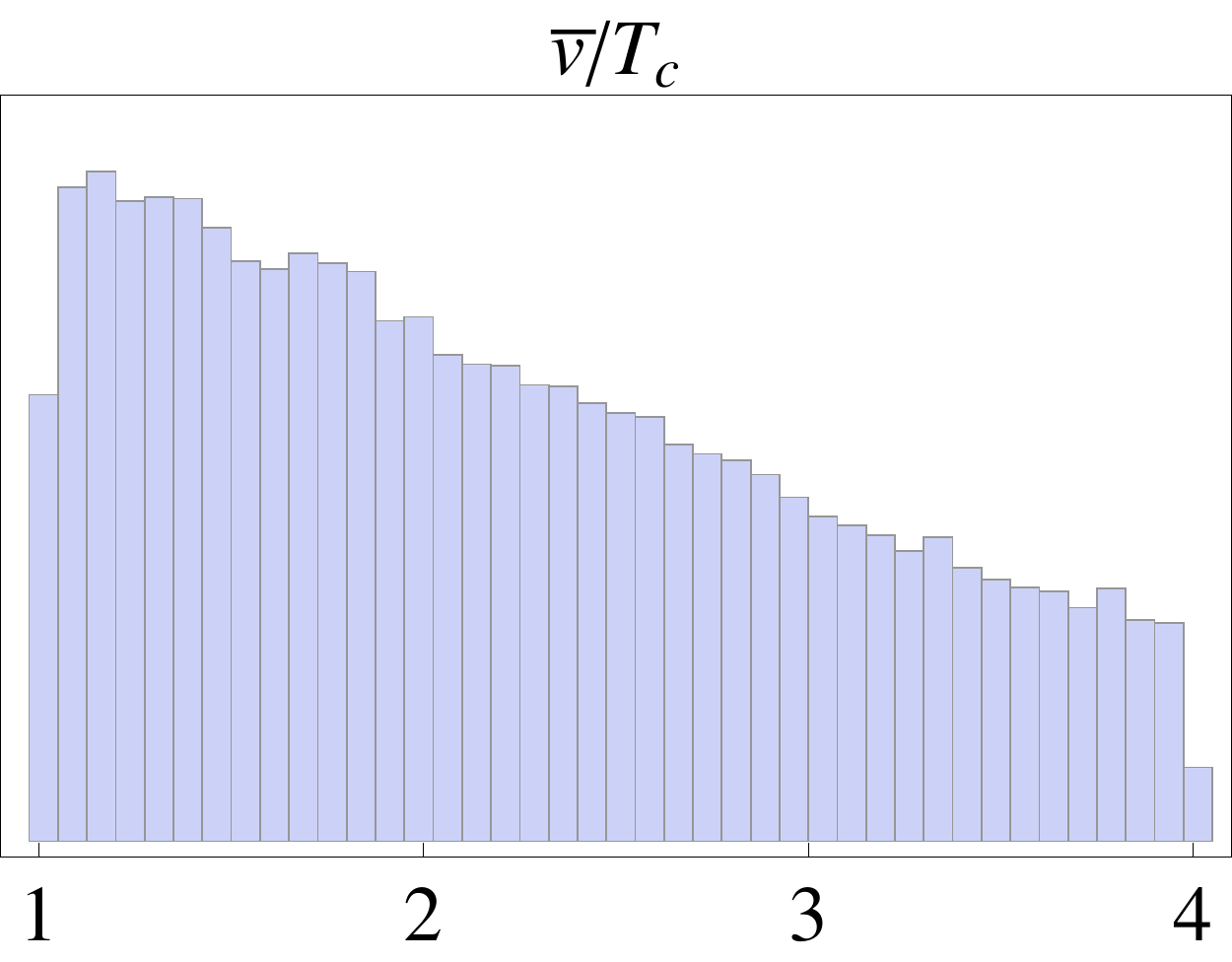}
\captionof{figure}{\it Distribution of the order parameter, $\bar{v}/T_c$, for models satisying all constraints. We find many models with $\bar{v}/T_c \gg 1$ indicating a strong electroweak phase transition as required for electroweak baryogenesis.}\label{fig:vc_Tc}
\end{Figure}

\begin{Figure}
\centering
\includegraphics[width=\linewidth]{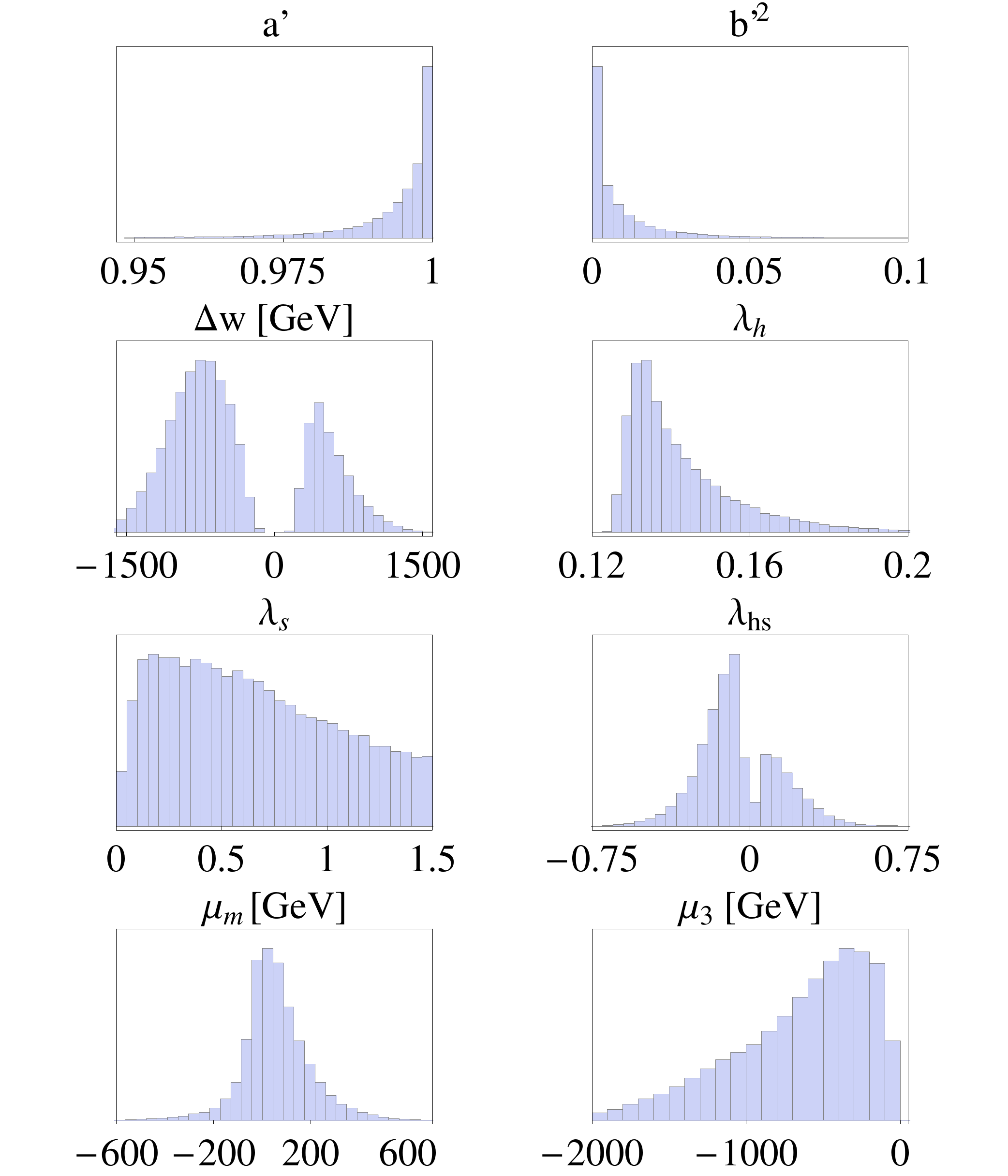}
\captionof{figure}{\it Distribution of some key parameters that satisfy all constraints from dark matter, the electroweak phase transition, Higgs physics, and electroweak precision tests.}\label{fig: params}
\end{Figure}

\section{Conclusion}
We have considered a minimal extension of the SM and showed that it can simultaneously explain DM while providing a strongly first order electroweak phase transition as required for electroweak baryogenesis. In contrast with some recent attempts in this area we find that by considering terms linear and cubic in $s$ and adding a singlet fermion both problems are easily solved. It is also possible to avoid current constraints from LHC Higgs physics because small mixing angles are preferred.

The next generation of dark matter direct detection experiments could push the upper bound on $\sigma_{SI}$ to $\sim 10^{-47}$ cm$^2$ \cite{Xenon1T}. This would rule out a significant proportion of the parameter space but very small values of $\sin \alpha$ allow will allow these limits to be evaded for $m_{\psi} \sim \frac{1}{2} m_{h_2}$ or $m_{\psi} > m_{h_2}$. If future LHC data improves the Higgs physics constraints from the 10\% level to the 1\% level this would still not be enough to rule out this model completely (see Figure \ref{fig:direct detection 2}).

\begin{Figure}
\centering
\includegraphics[width=.7\linewidth ,height=.5\linewidth]{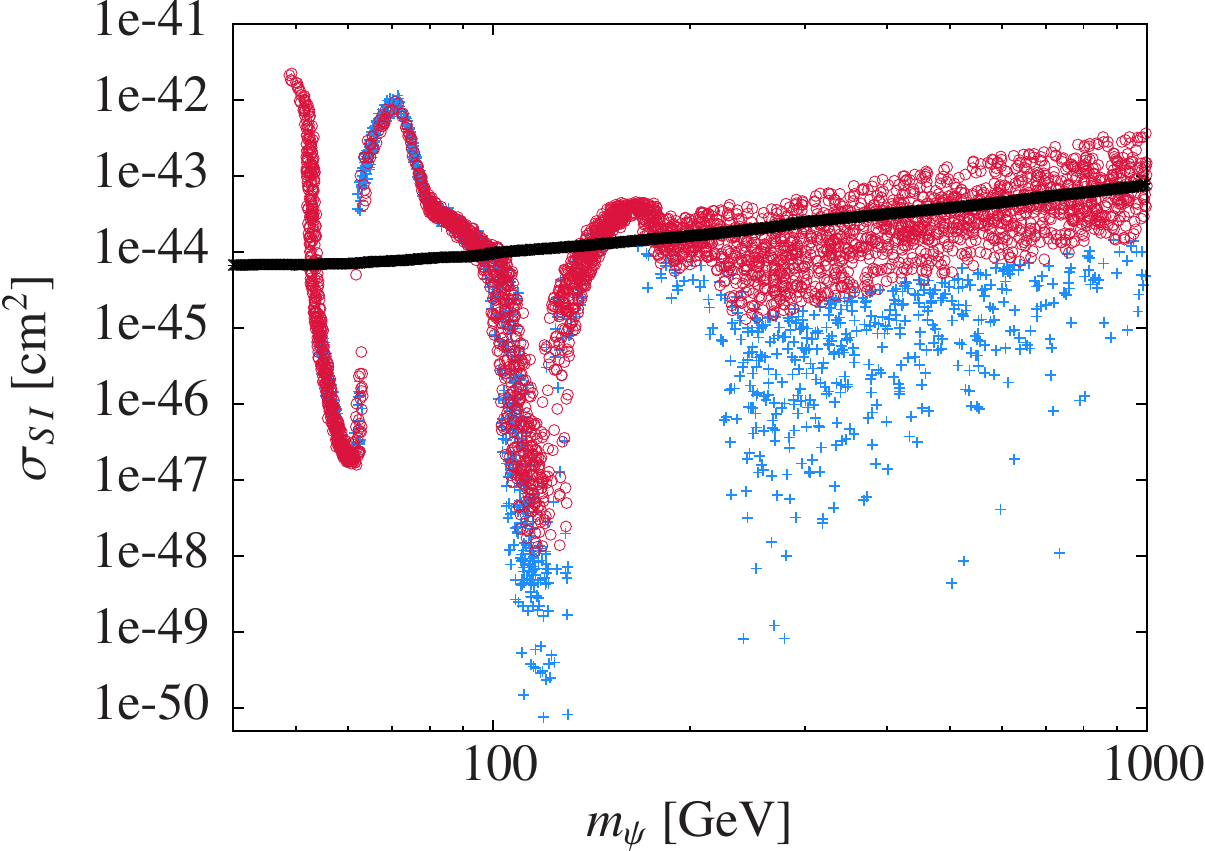}
\captionof{figure}{\it Scattering cross section per nucleon for $m_{h_2} = 250$ GeV $(\pm 5 \%)$ with the constraints on $a'$ and $b'^2$ improved from 10\% level to 1\% level. The Xenon100 2012 bound is included for comparision.}
\label{fig:direct detection 2}
\end{Figure}

It light of the recent interest in the stability of the electroweak vacuum to high scales\cite{Holthausen1112,EliasMiro1112,Xing1112,Rodejohann1203,Bezrukov1205, Degrassi1205,Alekhin1207,Masina1209,Chao1210} we note that models with additional scalars coupling to the Higgs can easily solve this problem   \cite{Gonderinger0910,Profumo1009,Chen1202,Gonderinger1202,Cheung1203,
EliasMiro1203,Lebedev1203,Frigerio1204,Baek1209,Abada1304,Barroso1303}.

It has been argued \cite{Profumo1009} for the case of scalar singlet dark matter that constraints from indirect detection of dark matter via annihilation into gamma rays \cite{FermiLAT1001} can rule out the region of resonant annihilation. In the case of this model the analogous amplitude for annihilation into gamma rays would be suppressed by mixing and a mass scale and so would be considerably weakened. It would nevertheless be interesting to consider the impact of indirect detection on the regions of our parameter space that are most resilient to constraints from direct detection experiments.

\section*{Acknowledgements}
The authors would like to thank Tevong You for discussions, and we are grateful to Jose Ramon Espinosa, Francesco Riva, Thomas Konstandin, and Laura Lopez-Honorez for commenting on the manuscript.

\section*{Appendix}
The cross section for $\psi \bar{\psi} \rightarrow h_i h_j$ is given by $s-$, $t-$, and $u-$channel terms, and the interference term\cite{Qin11}, 
\begin{equation*}
\sigma v_{rel}^{ij,s}=\kappa (s-4m_{\psi}^2) \left| \displaystyle\sum\limits_{r=1,2}\frac{g_r \lambda_r^{ij} }{s-m_{h_r}^2 +i m_{h_r} \Gamma_{h_r} } \right|^2,
\end{equation*}
\
\begin{equation*}
\begin{split}
\sigma v_{rel}^{ij,t/u}=& 2\kappa g_i^2g_j^2  \Biggl[  \frac{(4m_{\psi}^2 -m_{h_i})(4m_{\psi}^2 -m_{h_j})}{B^2 -A^2}    \\
 &  - \ln \left| \frac{A+B}{A-B} \right| \bigg( \frac{s+8m_{\psi}^2 -m_{h_i}^2-m_{h_i}^2}{2B} \\
 &  +\frac{16m_{\psi}^4-4s m_{\psi}^2-m_{h_i}^2m_{h_j}^2}{AB}  -2 \bigg)  \Biggr],
 \end{split}
\end{equation*}

\begin{equation*}
\begin{split}
\sigma v_{rel}^{ij,int} = &  4 \kappa g_i g_j m_{\psi}  \left(\displaystyle\sum\limits_{r=1,2}\frac{g_r \lambda_r^{ij}(s-m_{h_r}^2) }{(s-m_{h_r}^2)^2 + m_{h_r}^2 \Gamma_{h_r}^2 } \right) \\
  \times & \left( {\ln \left| \frac{A+B}{A-B} \right| {\left( \frac{A}{B} +\frac{1}{2}\frac{\beta_{\psi}}{\beta_{ij}}  -2\right) } }\right),
 \end{split}
\end{equation*}
where
\begin{equation*}
\kappa =\frac{1}{16 \pi s} \sqrt{\beta_{ij}^2+\frac{4 m_{h_i}m_{h_j}(m_{h_i}m_{h_j}-1)}{s^2}},
\end{equation*}

\begin{equation*}
A=\frac{1}{2}(m_{h_i}^2+m_{h_j}^2-s),\hspace{0.5 cm} D=\frac{s}{2}\beta_{\psi}\beta_{ij},
\end{equation*}
and
\begin{equation*}
\beta_{ij}=\sqrt{1-\frac{(m_{h_i}+m_{h_j})^2}{s}}\sqrt{1-\frac{(m_{h_i}-m_{h_j})^2}{s}}.
\end{equation*}
There is an additional symmetry factor of $\frac{1}{2}$ when $i=j$.

The cross-section for SM finals states is given by  \cite{Kim0803}
\begin{equation*}
\begin{split}
\sigma v_{rel}=&\frac{(g_s \sin \alpha \cos \alpha)^2}{16 \pi} \beta_{\psi}^2 \left| \displaystyle\sum\limits_{r=1,2}\frac{g_r \lambda_r^{ij} }{s-m_{h_r}^2 +i m_{h_r} \Gamma_{h_r} } \right|^2 \\
&\times \Biggl(  \displaystyle\sum\limits_{V=W,Z}  \epsilon \left( {\frac{2 m_V^2}{v}} \right)^2 \left( {2+ \frac{(s-2m_V^2)^2}{4m_V^4} }\right) \beta_V  \\
&  +\displaystyle\sum\limits_{q=b,t} N_c y_q s \beta_q^{3/2}  \Biggr) ,
\end{split}
\end{equation*}
where $N_c=3$ is the number of colors, $y_q$ are the quark Yukawa couplings, $\beta_i =\sqrt{1- 4m_i^2/s}$, and $\epsilon =1/2$ for $V=Z$ and otherwise unity. In the above the Standard Model contribution to $\Gamma_{h_i}$ was calculated using the HDECAY code \cite{Djouadi9704}.

The couplings for the cross sections are given by
\begin{align*}
g_r= & \left\{
     \begin{array}{lr}
       g_s \sin \alpha, & \text{if } i=1\\
       g_s \cos \alpha, & \text{if } i=2
     \end{array}
   \right. \\
 \lambda_{1}^{12}  =& c^3 \left(\frac{\mu_m}{4}+\frac{\lambda_{hs} w}{2}\right) -c^2   s   \left(3 \lambda_h v +\lambda_{hs} v \right) \\ 
&+   s^2 c  \  \left( \mu_3 -\frac{\mu_m}{2} -\lambda_{hs} w +3 \lambda_s w \right)- s^3  \  \frac{1}{2} \lambda_{hs} v , \\
\lambda^{12}_2  =& s^3\left(\frac{\mu_m}{4}+\frac{\lambda_{hs} w}{2}\right) + s^2 c   \left(3 \lambda_h v +\lambda_{hs} v\right) \\ 
&+ c^2 s  \  \left( \mu_3 -\frac{\mu_m}{2} -\lambda_{hs} w +3 \lambda_s w \right)+c^3  \  \frac{1}{2} \lambda_{hs} v, \\
 \lambda^{11}_1  =&\ c^3    \lambda_h v +c^2   s   \left(\frac{\mu_m}{4}+\frac{\lambda_{hs} w}{2}\right)\\
  &+ s^2 c \frac{\lambda_{hs} v}{2}+s^3 \left( \frac{\mu_3}{3}+\lambda_s w \right), \\
\lambda^{22}_2  =& -s^3  \  \lambda_h v + s^2 c   \left(\frac{\mu_m}{4}+\frac{\lambda_{hs} w}{2}\right) \\ 
&- c^2 s \  \frac{\lambda_{hs} v}{2}+ c^3  \ \left( \frac{\mu_3}{3}+\lambda_s w \right),
\end{align*}
where $c=\cos \alpha$ and $s= \sin \alpha$.
\vspace{0.5 cm}
\hrule

\bibliographystyle{h-physrev}
\bibliography{bibliography}

\begin{thebibliography}{10}

\bibitem{ATLAS1207}
ATLAS Collaboration, G.~Aad {\em et~al.},
\newblock Phys.Lett. {\bf B716}, 1 (2012), 1207.7214.

\bibitem{CMS1207}
CMS Collaboration, S.~Chatrchyan {\em et~al.},
\newblock Phys.Lett. {\bf B716}, 30 (2012), 1207.7235.

\bibitem{McDonald0702}
J.~McDonald,
\newblock Phys.Rev. {\bf D50}, 3637 (1994), hep-ph/0702143.

\bibitem{Davoudiasl0405}
H.~Davoudiasl, R.~Kitano, T.~Li, and H.~Murayama,
\newblock Phys.Lett. {\bf B609}, 117 (2005), hep-ph/0405097.

\bibitem{Burgess0011}
C.~Burgess, M.~Pospelov, and T.~ter Veldhuis,
\newblock Nucl.Phys. {\bf B619}, 709 (2001), hep-ph/0011335.

\bibitem{O'Connell0611}
D.~O'Connell, M.~J. Ramsey-Musolf, and M.~B. Wise,
\newblock Phys.Rev. {\bf D75}, 037701 (2007), hep-ph/0611014.

\bibitem{BahatTreidel0611}
O.~Bahat-Treidel, Y.~Grossman, and Y.~Rozen,
\newblock JHEP {\bf 0705}, 022 (2007), hep-ph/0611162.

\bibitem{Barger0706}
V.~Barger, P.~Langacker, M.~McCaskey, M.~J. Ramsey-Musolf, and G.~Shaughnessy,
\newblock Phys.Rev. {\bf D77}, 035005 (2008), 0706.4311.

\bibitem{He0701}
X.-G. He, T.~Li, X.-Q. Li, and H.-C. Tsai,
\newblock Mod.Phys.Lett. {\bf A22}, 2121 (2007), hep-ph/0701156.

\bibitem{Gonderinger1202}
M.~Gonderinger, H.~Lim, and M.~J. Ramsey-Musolf,
\newblock Phys.Rev. {\bf D86}, 043511 (2012), 1202.1316.

\bibitem{Kim0611}
Y.~G. Kim and K.~Y. Lee,
\newblock Phys.Rev. {\bf D75}, 115012 (2007), hep-ph/0611069.

\bibitem{Kim0803}
Y.~G. Kim, K.~Y. Lee, and S.~Shin,
\newblock JHEP {\bf 0805}, 100 (2008), 0803.2932.

\bibitem{Qin11}
H.-Y. Qin, W.-Y. Wang, and Z.-H. Xiong,
\newblock Chin.Phys.Lett. {\bf 28}, 111202 (2011).

\bibitem{LopezHonorez1203}
L.~Lopez-Honorez, T.~Schwetz, and J.~Zupan,
\newblock Phys.Lett. {\bf B716}, 179 (2012), 1203.2064.

\bibitem{Baek1209}
S.~Baek, P.~Ko, W.-I. Park, and E.~Senaha,
\newblock JHEP {\bf 1211}, 116 (2012), 1209.4163.

\bibitem{Farina1303}
M.~Farina, D.~Pappadopulo, and A.~Strumia,
\newblock (2013), 1303.7244.

\bibitem{Petraki0711}
K.~Petraki and A.~Kusenko,
\newblock Phys.Rev. {\bf D77}, 065014 (2008), 0711.4646.

\bibitem{Sakharov1967}
A.~Sakharov,
\newblock Pisma Zh.Eksp.Teor.Fiz. {\bf 5}, 32 (1967).

\bibitem{Gavela9312}
M.~Gavela, P.~Hernandez, J.~Orloff, and O.~Pene,
\newblock Mod.Phys.Lett. {\bf A9}, 795 (1994), hep-ph/9312215.

\bibitem{Gavela9406}
M.~Gavela, P.~Hernandez, J.~Orloff, O.~Pene, and C.~Quimbay,
\newblock Nucl.Phys. {\bf B430}, 382 (1994), hep-ph/9406289.

\bibitem{Huet9404}
P.~Huet and E.~Sather,
\newblock Phys.Rev. {\bf D51}, 379 (1995), hep-ph/9404302.

\bibitem{Kuzmin85}
V.~Kuzmin, V.~Rubakov, and M.~Shaposhnikov,
\newblock Phys.Lett. {\bf B155}, 36 (1985).

\bibitem{Cohen9302}
A.~G. Cohen, D.~Kaplan, and A.~Nelson,
\newblock Ann.Rev.Nucl.Part.Sci. {\bf 43}, 27 (1993), hep-ph/9302210.

\bibitem{Trodden9803}
M.~Trodden,
\newblock Rev.Mod.Phys. {\bf 71}, 1463 (1999), hep-ph/9803479.

\bibitem{Morrissey1206}
D.~E. Morrissey and M.~J. Ramsey-Musolf,
\newblock New J.Phys. {\bf 14}, 125003 (2012), 1206.2942.

\bibitem{Pietroni9207}
M.~Pietroni,
\newblock Nucl.Phys. {\bf B402}, 27 (1993), hep-ph/9207227.

\bibitem{Menon0404}
A.~Menon, D.~Morrissey, and C.~Wagner,
\newblock Phys.Rev. {\bf D70}, 035005 (2004), hep-ph/0404184.

\bibitem{Profumo0705}
S.~Profumo, M.~J. Ramsey-Musolf, and G.~Shaughnessy,
\newblock JHEP {\bf 0708}, 010 (2007), 0705.2425.

\bibitem{Cline0905}
J.~M. Cline, G.~Laporte, H.~Yamashita, and S.~Kraml,
\newblock JHEP {\bf 0907}, 040 (2009), 0905.2559.

\bibitem{Espinosa1107}
J.~R. Espinosa, T.~Konstandin, and F.~Riva,
\newblock Nucl.Phys. {\bf B854}, 592 (2012), 1107.5441.

\bibitem{Chung1209}
D.~J. Chung, A.~J. Long, and L.-T. Wang,
\newblock Phys.Rev. {\bf D87}, 023509 (2013), 1209.1819.

\bibitem{Cline1210}
J.~M. Cline and K.~Kainulainen,
\newblock JCAP {\bf 1301}, 012 (2013), 1210.4196.

\bibitem{Patel1212}
H.~H. Patel and M.~J. Ramsey-Musolf,
\newblock (2012), 1212.5652.

\bibitem{Espinosa9301}
J.~Espinosa and M.~Quiros,
\newblock Phys.Lett. {\bf B305}, 98 (1993), hep-ph/9301285.

\bibitem{Choi9308}
J.~Choi and R.~Volkas,
\newblock Phys.Lett. {\bf B317}, 385 (1993), hep-ph/9308234.

\bibitem{Ham0411}
S.~Ham, Y.~Jeong, and S.~Oh,
\newblock J.Phys. {\bf G31}, 857 (2005), hep-ph/0411352.

\bibitem{Ahriche0701}
A.~Ahriche,
\newblock Phys.Rev. {\bf D75}, 083522 (2007), hep-ph/0701192.

\bibitem{Espinosa1110}
J.~R. Espinosa, B.~Gripaios, T.~Konstandin, and F.~Riva,
\newblock JCAP {\bf 1201}, 012 (2012), 1110.2876.

\bibitem{Barger0811}
V.~Barger, P.~Langacker, M.~McCaskey, M.~Ramsey-Musolf, and G.~Shaughnessy,
\newblock Phys.Rev. {\bf D79}, 015018 (2009), 0811.0393.

\bibitem{Chowdhury1110}
T.~A. Chowdhury, M.~Nemevsek, G.~Senjanovic, and Y.~Zhang,
\newblock JCAP {\bf 1202}, 029 (2012), 1110.5334.

\bibitem{Ahriche1201}
A.~Ahriche and S.~Nasri,
\newblock Phys.Rev. {\bf D85}, 093007 (2012), 1201.4614.

\bibitem{Cline1302}
J.~M. Cline and K.~Kainulainen,
\newblock (2013), 1302.2614.

\bibitem{Frigerio1204}
M.~Frigerio, A.~Pomarol, F.~Riva, and A.~Urbano,
\newblock JHEP {\bf 1207}, 015 (2012), 1204.2808.

\bibitem{ATLAS-CONF-2012-163}
CERN Report No. ATLAS-CONF-2012-163, 2012 (unpublished).

\bibitem{Espinosa1205}
J.~R. Espinosa, M.~Muhlleitner, C.~Grojean, and M.~Trott,
\newblock JHEP {\bf 1209}, 126 (2012), 1205.6790.

\bibitem{Ellis1303}
J.~Ellis and T.~You,
\newblock (2013), 1303.3879.

\bibitem{Falkowski1303}
A.~Falkowski, F.~Riva, and A.~Urbano,
\newblock (2013), 1303.1812.

\bibitem{CMS1304}
CMS Collaboration, S.~Chatrchyan {\em et~al.},
\newblock (2013), 1304.0213.

\bibitem{Baek1112}
S.~Baek, P.~Ko, and W.-I. Park,
\newblock JHEP {\bf 1202}, 047 (2012), 1112.1847.

\bibitem{Baak1209}
M.~Baak {\em et~al.},
\newblock Eur.Phys.J. {\bf C72}, 2205 (2012), 1209.2716.

\bibitem{Eberhardt1209}
O.~Eberhardt {\em et~al.},
\newblock Phys.Rev.Lett. {\bf 109}, 241802 (2012), 1209.1101.

\bibitem{PlanckXVI}
Planck Collaboration, P.~Ade {\em et~al.},
\newblock (2013), 1303.5076.

\bibitem{Gondolo1990}
P.~Gondolo and G.~Gelmini,
\newblock Nucl.Phys. {\bf B360}, 145 (1991).

\bibitem{Sommerfield31}
A.~Sommerfeld,
\newblock Annalen der Physik {\bf 403}, 257 (1931).

\bibitem{Nihei0404}
T.~Nihei and M.~Sasagawa,
\newblock Phys.Rev. {\bf D70}, 055011 (2004), hep-ph/0404100.

\bibitem{Ellis0001}
J.~R. Ellis, A.~Ferstl, and K.~A. Olive,
\newblock Phys.Lett. {\bf B481}, 304 (2000), hep-ph/0001005.

\bibitem{Bottino9612}
A.~Bottino {\em et~al.},
\newblock Phys.Lett. {\bf B402}, 113 (1997), hep-ph/9612451.

\bibitem{Xenon1207}
XENON100 Collaboration, E.~Aprile {\em et~al.},
\newblock Phys.Rev.Lett. {\bf 109}, 181301 (2012), 1207.5988.

\bibitem{Xenon1T}
XENON1T collaboration, E.~Aprile,
\newblock (2012), 1206.6288.

\bibitem{Carson90}
L.~Carson, X.~Li, L.~McLerran, and R.-T. Wang,
\newblock Phys. Rev. D {\bf 42}, 2127 (1990).

\bibitem{Patel1101}
H.~H. Patel and M.~J. Ramsey-Musolf,
\newblock JHEP {\bf 1107}, 029 (2011), 1101.4665.

\bibitem{Wainwright1204}
C.~L. Wainwright, S.~Profumo, and M.~J. Ramsey-Musolf,
\newblock Phys.Rev. {\bf D86}, 083537 (2012), 1204.5464.

\bibitem{Espinosa0809}
J.~Espinosa, T.~Konstandin, J.~No, and M.~Quiros,
\newblock Phys.Rev. {\bf D78}, 123528 (2008), 0809.3215.

\bibitem{Holthausen1112}
M.~Holthausen, K.~S. Lim, and M.~Lindner,
\newblock JHEP {\bf 1202}, 037 (2012), 1112.2415.

\bibitem{EliasMiro1112}
J.~Elias-Miro {\em et~al.},
\newblock Phys.Lett. {\bf B709}, 222 (2012), 1112.3022.

\bibitem{Xing1112}
Z.-z. Xing, H.~Zhang, and S.~Zhou,
\newblock Phys.Rev. {\bf D86}, 013013 (2012), 1112.3112.

\bibitem{Rodejohann1203}
W.~Rodejohann and H.~Zhang,
\newblock JHEP {\bf 1206}, 022 (2012), 1203.3825.

\bibitem{Bezrukov1205}
F.~Bezrukov, M.~Y. Kalmykov, B.~A. Kniehl, and M.~Shaposhnikov,
\newblock JHEP {\bf 1210}, 140 (2012), 1205.2893.

\bibitem{Degrassi1205}
G.~Degrassi {\em et~al.},
\newblock JHEP {\bf 1208}, 098 (2012), 1205.6497.

\bibitem{Alekhin1207}
S.~Alekhin, A.~Djouadi, and S.~Moch,
\newblock Phys.Lett. {\bf B716}, 214 (2012), 1207.0980.

\bibitem{Masina1209}
I.~Masina,
\newblock Phys.Rev. {\bf D87}, 053001 (2013), 1209.0393.

\bibitem{Chao1210}
W.~Chao, M.~Gonderinger, and M.~J. Ramsey-Musolf,
\newblock Phys.Rev. {\bf D86}, 113017 (2012), 1210.0491.

\bibitem{Gonderinger0910}
M.~Gonderinger, Y.~Li, H.~Patel, and M.~J. Ramsey-Musolf,
\newblock JHEP {\bf 1001}, 053 (2010), 0910.3167.

\bibitem{Profumo1009}
S.~Profumo, L.~Ubaldi, and C.~Wainwright,
\newblock Phys.Rev. {\bf D82}, 123514 (2010), 1009.5377.

\bibitem{Chen1202}
C.-S. Chen and Y.~Tang,
\newblock JHEP {\bf 1204}, 019 (2012), 1202.5717.

\bibitem{Cheung1203}
C.~Cheung, M.~Papucci, and K.~M. Zurek,
\newblock JHEP {\bf 1207}, 105 (2012), 1203.5106.

\bibitem{EliasMiro1203}
J.~Elias-Miro, J.~R. Espinosa, G.~F. Giudice, H.~M. Lee, and A.~Strumia,
\newblock JHEP {\bf 1206}, 031 (2012), 1203.0237.

\bibitem{Lebedev1203}
O.~Lebedev,
\newblock Eur.Phys.J. {\bf C72}, 2058 (2012), 1203.0156.

\bibitem{Abada1304}
A.~Abada and S.~Nasri,
\newblock (2013), 1304.3917.

\bibitem{Barroso1303}
A.~Barroso, P.~Ferreira, I.~Ivanov, and R.~Santos,
\newblock JHEP {\bf 1306}, 045 (2013), 1303.5098.

\bibitem{FermiLAT1001}
A.~Abdo {\em et~al.},
\newblock Phys.Rev.Lett. {\bf 104}, 091302 (2010), 1001.4836.

\bibitem{Djouadi9704}
A.~Djouadi, J.~Kalinowski, and M.~Spira,
\newblock Comput.Phys.Commun. {\bf 108}, 56 (1998), hep-ph/9704448.

\end{thebibliography}

\end{multicols}
\end{document}